\documentstyle[prl,twocolumn,aps,graphicx]{revtex}

\begin{document}

\draft

\title{Magneto infra-red absorption in high electronic density  GaAs quantum wells}

\author{A. J. L. Poulter, J. Zeman\cite{byline},
          D. K. Maude, M. Potemski and G. Martinez}
\address{ Grenoble High Magnetic Field Laboratory MPI-FKF/CNRS,
         25 Avenue des Martyrs, F-38042 Grenoble Cedex 9, France}
\author{A. Riedel, R. Hey and K. J. Friedland }
\address{ Paul Drude Institute, Hausvogteiplatz 5-7, D-10117 Berlin,
          Germany }

\date{\today}
\maketitle

\begin{abstract}
Magneto infra-red absorption measurements have been performed in
a highly doped GaAs quantum well which has been lifted off and
bonded to a silicon substrate, in order to study the resonant
polaron interaction. It is found that the pinning of the
cyclotron energy occurs at an energy close to that of the
transverse optical phonon of GaAs. This unexpected result is
explained by a model taking into account the full dielectric
constant of the quantum well.
\end{abstract}
\vskip 5mm The Fr\"{o}hlich interaction is one of the main
electron-phonon intrinsic interactions in polar materials, and
has attracted much attention both in 3-dimensional (3D) [1] and
2-dimensional (2D) [2, 3] structures. The most spectacular
manifestation of this interaction is the effect of resonant
magneto-polaron coupling (RMPC), i.e. an anticipated anticrossing
behaviour between the $\vert n=0 + {\rm \, 1 \, LO \, phonon}
\rangle$ state  and the $\vert n=1 \rangle$ state, $n$ being the
Landau level (LL) index, when the cyclotron frequency
$\omega_{c}=eB/m^{*}$ equals the longitudinal optical phonon
frequency $\omega_{LO}$ ($B$ is the applied magnetic field and
$m^{*}$ the carrier effective mass). This has been studied
theoretically using both perturbation theories [1, 2] and the
memory function approach [3]. Whereas the former method
restricted the analysis to vanishing doping levels, the latter
has been extended to include the electron-electron interaction,
predicting a disappearance of the coupling with increasing
electron density. However in doped systems $\omega_{LO}$ is no
longer a normal mode of the system and couples to the plasma mode
$\omega_{p}$ in 3D [4] or to intersubband plasmon modes in 2D [5,
6], thus giving rise to new longitudinal modes with frequencies
$\omega_{L}^{-}$ and $\omega_{L}^{+}$. We argue here that these
new coupled modes are the only excitations which can couple to
electrons via their associated macrosopic electric field.
Therefore the theoretical treatment of doped systems assuming the
existence of uncoupled pure LO phonons is obviously not
consistent.

Experimentally there have been numerous investigations into
resonant polaronic coupling [7--11], but the results are often
obscured by the presence of the reststrahlen band, and the strong
dielectric effects in this region make the comparison between
experiment and theory unreliable. This is a major obstacle that
we have been able to overcome in the present investigation. In
this letter we report on polaronic effects in a high carrier
density GaAs quantum well. The experiment shows no polaron
coupling involving the undressed LO phonon mode, instead the
pinning of the cyclotron resonance at a frequency close to
$\omega_{TO}$ is clearly observed. The pinning energy is explained
in terms of coupling of the cyclotron resonance excitation with
the magneto-plasmon-LO-phonon mode $\omega_{L}^{-}$ which occurs
in this structure around the frequency of the transverse optical
(TO) phonon $\omega_{TO}$.

The samples studied were grown by MBE on semi-insulating GaAs
substrates as described in Ref. 12. A quantum well (QW), of width
$L=10$~nm, is sandwiched between two 60 period AlAs/GaAs
superlattices which are $\delta$-doped by silicon in the third
GaAs well on either side of the QW. This results in high mobility
electrons in the QW, with additional electrons being trapped by
AlAs X-band-like states in the barrier layer adjacent to the doped
GaAs well. This configuration provides high electron densities,
high electron mobilities and also maintains a symmetric band
structure. A number of such samples have been studied
demonstrating all nonlinear effects in the RMPC region but the
data at higher fields are obscured by the reststrahlen band
absorption. In order to overcome this problem we have, for one of
the samples, lifted off the epi-layer from the substrate by
selective etching and deposited it on a wedged silicon substrate.
The sample area is of the order of 5~mm$^{2}$ and Hall
measurements on a parent sample provided a value for the carrier
concentration $n_{s} = 1.28 \times 10^{12}$~cm$^{-2}$ and a
mobility of 114~m$^{2}$/Vs. Band structure calculations show that
the splitting $E_{01}$ of the two lowest electric subbands
$E_{0}$ and $E_{1}$ equals 120~meV and the 2D electron gas (2DEG)
occupy only the $E_{0}$ level. Transmission measurements on this
sample were performed with a Bruker IFS--113 Fourier transform
spectrometer, at a temperature of 1.7~K, in magnetic fields $B$
up to 28~T applied perpendicularly to the 2DEG plane. The infrared
radiation was detected by a silicon bolometer placed in situ
behind the sample which was mounted on a rotating sample holder.
For each value of $B$ a reference and sample spectra are recorded
and ratioed, the reference spectrum being that of a piece of the
wedged silicon substrate. Therefore the results displayed in
Fig.~1 are the absolute transmission spectra eliminating the
response of the experimental setup.

The spectra (Fig. 1) exhibit two well defined TO absorption lines
at $\hbar \omega_{TO}{\rm (GaAs)} = 33.6$~meV and $\hbar
\omega_{TO}{\rm (AlAs)} = 44.9$~meV together with a
characteristic $B$ dependent cyclotron resonance (CR) absorption
line which appears as a single line at low and high fields, but
splits into two components below and nearby $\omega_{TO}$(GaAs)
(see below). It is important to note (Fig.~1b) that the
observation of the CR structure is obscured only in a narrow
energy range around $\omega_{TO}$(GaAs) but not in the region
where the CR line coincides with $\omega_{LO}$(GaAs). To
understand the data we have calculated the optical transmission
of the whole structure using the multi-layer dielectric model
[13] and the exact layer sequence of the samples. As evidenced in
Fig.~1, we obtain a very reasonable fit for both TO absorption
lines with the following high frequency dielectric constant,
energies and damping parameters $\varepsilon_{\infty}=
10.6$~(8.4), $\hbar \omega_{LO}= 36.3$~(49.1) and $\gamma_{TO}
=0.25$~(0.56) meV for GaAs (AlAs) respectively. The CR absorption
is mimicked neglecting occupation effects and polaronic coupling,
with a single oscillator corresponding to the reduced effective
mass $m^{*}/m_{0}=0.077$ (where $m_{0}$ is the free electron
mass) and a damping parameter $\gamma_{e} = 0.1$~meV.
\begin{figure}[h]
\begin{center}
\includegraphics[width=8cm,clip=true]{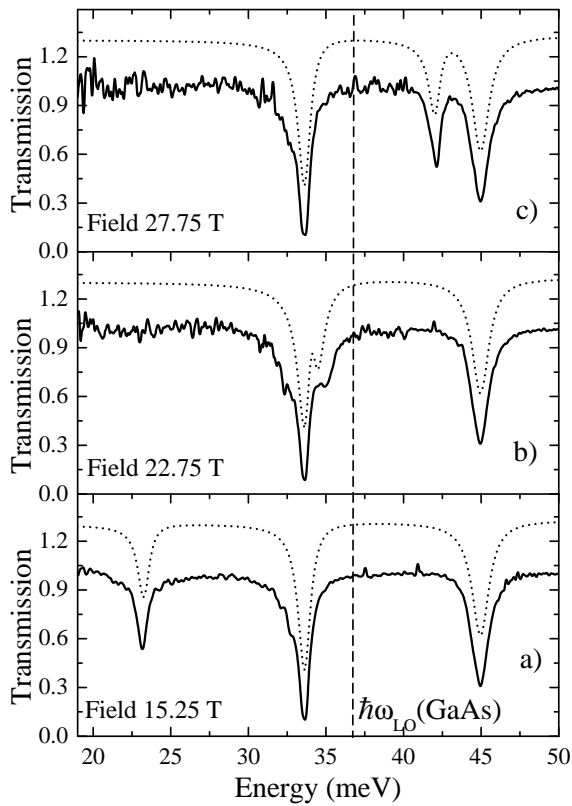}
\end{center}
\label{fig1} \caption{Examples of the spectra obtained at a)
15.25~T, b) 22.75~T and c) 27.75~T. The dotted lines show the
theoretical results (shifted vertically for clarity) of the
multi-layer transmission response of the structure as described
in the text.}
\end{figure}
Although we have been able to closely reproduce the experimental
results, the small low energy asymmetry of the GaAs TO line
(observed at any fields and possibly resulting from the confined
phonon models) is not accounted by the model. In order to be more
accurate we ratioed the transmission spectra at finite $B$ by the
absolute transmission spectra recorded at $B=0$. The resulting
curves are shown in Fig.~2. The $\omega_{TO}$ absorption is now
also absent and we can more clearly follow the evolution of the CR
absorption structure. Nevertheless, we prefer to be rather
conservative with respect to our data treatment and have ignored
structures appearing in the frequency range around $\hbar
\omega_{TO}$ when fitting the resonances with Lorentzians and
plotting, in Fig.~3, the obtained transition energies versus $B$.
Beyond 17~T when the filling factor $\nu < 3$ ($\nu= n_{s}/G_{B}$,
$G_{B}=eB/h$ is the LL degeneracy) the CR line clearly splits
into two components A and B (Figs.~2 and 3a) whereas a single CR
line denoted C is observed above 23~T.
\begin{figure}[h]
\begin{center}
\includegraphics[width=8cm,clip=true]{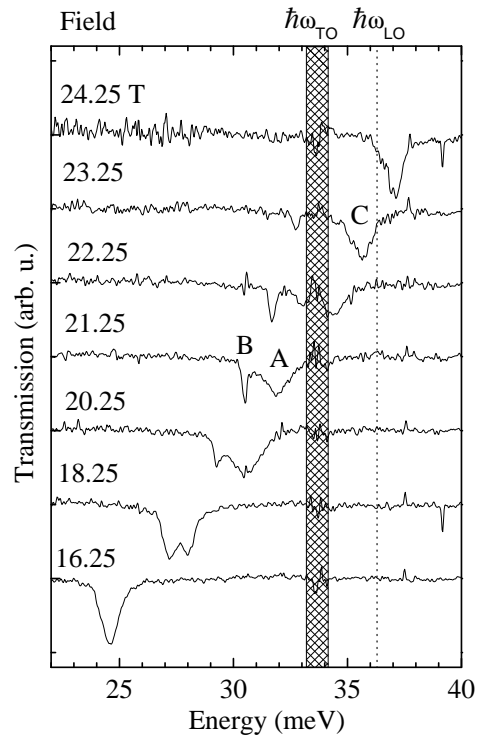}
\end{center}
\label{fig2} \caption{Characteristic spectra at various magnetic
fields. The splitting of the CR line into A and B components
appears clearly below $\hbar \omega_{TO}$. The hatched area
corresponds to the width of the TO absorption line.}
\end{figure}
Let us first concentrate on the two, A and B, CR components. The
line B is clearly observed for $\nu < 3$ (Fig.~3) and
progressively disappears from the spectrum. It is not observed in
high magnetic fields when $\nu$ approaches 2 because the $n=1$
LL gets depopulated. The splitting of
the CR into A and B components is attributed to two possible
different CR transitions involving the $n=0$ (line A) and the
second $n=1$ (line B) Landau level in the initial state (see
insert in Fig.~3). These two transitions involve opposite spin
electrons and can differ in energy due to the band
nonparabolicity (NP). Indeed, due to combined effects of
confinement, high doping levels and high magnetic fields one
spans an energy range where NP effects are important. In such a
case, one can write within the random phase approximation [5,
13], the generalized cyclotron active dielectric function
$\varepsilon_{a}$ of the GaAs in the Faraday configuration,
neglecting damping and polaronic effects as:
\begin{eqnarray}
\varepsilon_{a}(\omega) & = & \varepsilon_{\infty}
\frac{\omega_{LO}^{2}- \omega^{2}}{\omega_{TO}^{2}- \omega^{2}} -
\frac{4\pi e^{2} G_{B}}{L\omega} \times \nonumber \\
& &  \times \sum_{n=0}^{\infty} \sum_{\sigma} \frac{(f_{n,\sigma}
- f_{n+1,\sigma})(n+1)}{\left[ \omega- (E_{n+1, \sigma} - E_{n,
\sigma})/\hbar\right] m^{*}_{n, \sigma}},
\end{eqnarray}
where $f_{n,\sigma}$ is the distribution function of electrons in
the $n$-th LL, with spin $\sigma$, energy $E_{n,\sigma}$ and
effective masses $m^{*}_{n,\sigma}$ and $\sum_{n,\sigma} G_{B}(
f_{n,\sigma} - f_{n+1,\sigma})(n+1) =n_{s}$. Eq.~1 is the 3D
dielectric constant which is used for the QW layer in the
multi-layer dielectric model, the quasi 2D character being
introduced through the boundary conditions [13] imposed to the
electric and magnetic components of the radiation field at each
interface. Calculations performed with a 10-bands {\bf k.p} model
[14] reveal that the splitting between peaks A and B is due to
the NP effects with transitions originating from the $n=0$ and
$n=1$ LL respectively in accordance with the previous discussion
(Fig.~3a, thin lines).
\begin{figure}[h]
\begin{center}
 \includegraphics[width=74mm,clip=true]{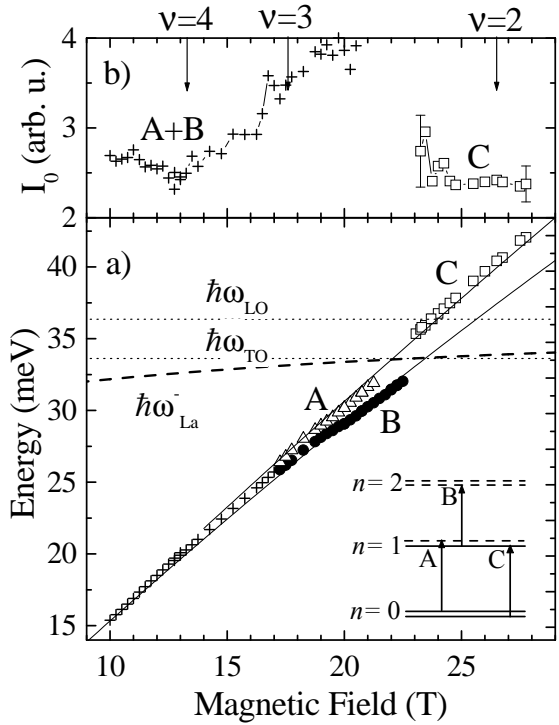}
\end{center}
\label{fig3} \caption{a) Magnetic field variation of the
different CR line energies, determined by fitting the zero field
ratioed data using Lorentzians. The thin solid lines correspond
to the calculated CR transitions including NP effects. b) The
corresponding total integrated absorption intensity $I_{0}$ in
arbitrary units for the CR lines.}
\end{figure}
It is natural to assume that lines A and C represent the same
transition observed on either side of $\hbar \omega_{TO}$. We
argue here that this transition suffers an interaction in the
vicinity of the $\hbar \omega_{TO}$. The interaction is clearly
inferred from the fact that the peaks A and C have a width
increasing significantly on both sides of $\hbar \omega_{TO}$
(Fig.~2). What convinces us more is the weak change in the slopes
of the energy variation of the lines A and C below and above
$\hbar \omega_{TO}$ and the increase of the \emph{total}
integrated CR intensity $I_{0}$ around $\hbar \omega_{TO}$
(Fig.~3b) whereas, as expected for a non interacting system,
$I_{0}$ remains constant at high and low fields. The amplitude of
an eventual anticrossing between lines A and C is , however, hard
to ascertain accurately due to possible experimental errors in
estimating the peak position in the closed vicinity of $\hbar
\omega_{TO}$. The interaction with phonon-like modes occurs
around $\hbar \omega_{TO}$ whereas no singularity is observed
around $\hbar \omega_{LO}$. The absence of this singularity has
already been reported [7,~8] on samples with lower doping levels
($\nu < 2$) but not explained.

An experiment that can also probe the electron-phonon like modes
is the magneto-phonon resonance (MPR) effect which manifests
itself as an oscillatory component, in reciprocal magnetic field,
of the resistivity. This component is proportional to $\cos (2\pi
\omega_{0}/\omega_{c})$ [15] where $\omega_{0}$ has been up till
now assumed to be equal to $\omega_{LO}$. We have performed
magneto-resistivity measurements on the parent sample and
observed the MPR oscillations between 5 and 15~T for temperatures
ranging from 90~K to 140~K (Fig.~4). These experiments provide the
mean value $\hbar \omega_{0}({\rm meV})m^{*}(m_{0}) = 2.60 \pm
0.05$ when interpreted with standard theory [15]. Taking for
$m^{*}$ the value of 0.078 measured by cyclotron resonance at this
temperature, we deduce a value $\hbar \omega_{0} = 33.3 \pm
0.7$~meV which is close to $\hbar \omega_{TO}$ and definitively
lower than $\hbar \omega_{LO}$. We therefore confirm that there
exist longitudinal modes with energies close to $\hbar
\omega_{TO}$ which interact with the CR mode.
\begin{figure}[h]
\begin{center}
\includegraphics[width=8cm,clip=true]{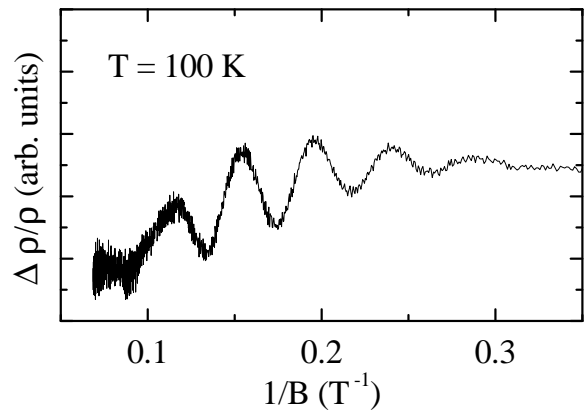}
\end{center}
\caption{Oscillatory component of the resistivity versus the
inverse of $B$.}
\end{figure}
In order to identify the nature of these modes, one can use a
simplified version of Eq.~1 neglecting the NP effects:
\begin{equation}
\varepsilon_{a}(\omega)  =  \varepsilon_{\infty}
\frac{\omega_{LO}^{2}- \omega^{2}}{\omega_{TO}^{2}- \omega^{2}} -
\frac{\omega^{2}_{p}}{\omega ( \omega- \omega_{c})}
\end{equation}
with $\omega_{p}^{2}=4\pi N e^{2}/m^{*}$, where $N=n_{s}/L$ is
the corresponding 3D electronic density. The longitudinal
solution of Eq.~2 ($\to \varepsilon_{a}=0$) are composed of two
modes $\omega_{La}^{-}$ and $\omega_{La}^{+}$ [16] starting from
$\omega_{L}^{-}$ and $\omega_{L}^{+}$ at $B=0$ and pinned to
$\omega= \omega_{LO}$ and $\omega=\omega_{c}$ respectively for
extreme values of $B$. It is clear from Eq.~2 that
$\varepsilon_{a}(\omega)$ has two poles at $\omega=\omega_{c}$
and $\omega=\omega_{TO}$ and therefore a zero (corresponding to
$\omega_{La}^{-}$) in between. When
$\omega=\omega_{c}=\omega_{TO}$, $\omega_{La}^{-}$ coincides with
the common pole. This property is independent on $\omega_{p}^{2}$
(and therefore on $N$); the $\omega_{La}^{-}(N,B)$ curves display
fixed point at $\omega=\omega_{TO}$. The resulting curve
$\omega_{La}^{-}(B)$ is displayed in Fig.~3a for the present
case: it is clear that the resonant interaction should occur
between \emph{this mode} and the CR mode always at
$\omega=\omega_{TO}$. The calculated mean frequency is in
reasonable agreement with our experimental findings. Though the
resonant interaction occurs at $\omega=\omega_{TO}$ this does not
imply that the maximum interaction occurs at this energy: the
reason is that the macroscopic electric field associated with
$\omega_{La}^{-}$ and which governs the strength of the
interaction goes to zero, in the absence of damping, when
$\omega_{La}^{-} = \omega_{TO}$. One can note, for instance, that
the amplitude of the MPR oscillations decreases at high fields
which could also be the signature of the vanishing interaction.

This interpretation is based on a 3D-like model for the
dielectric constant. In a 2D model [5, 6], intersubband plasmon
modes also exist. They can be measured by resonant Raman
scattering measurements [17] which in this sample give a
longitudinal transition at $35.5\pm 0.05$~meV. This mode which
propagates in a 2D plane, is in the infra-red range essentially
independent on $B$ [6]. It does not seem that the observed
coupling occurs with this mode. A more quantitative analysis
requires theoretical efforts for a complete treatment of polaronic
coupling with magneto-plasmon-phonon modes.

In conclusion, we have performed magneto-absorption measurements
on a highly doped epilayer lifted off the substrate and bonded to
a silicon substrate. We have demonstrated that near the
$\omega_{TO}$ frequency the CR line splits into two components
due to occupation and non-parabolicity effects. The CR transition
originating from the $n=0$ LL has been investigated through the
reststrahlen band and no coupling with undressed LO phonon modes
has been observed. The CR mode has been found to interact with a
longitudinal mode which has been identified by magneto-phonon
resonance experiment and which, for our sample, has an energy
close to $\hbar \omega_{TO}$. We have assigned this mode to the
low energy coupled magneto-plasmon-phonon mode of the GaAs QW. We
hope that this report will stimulate theoretical work to identify
the nature of the interaction between electrons and this type of
modes.

\acknowledgments The Grenoble High Magnetic Field Laboratory is
``Laboratoire conventionn\'{e} \`{a} l'UJF et l'INPG de
Grenoble''.  J. Z. acknowledges the partial supports from the
grant ERBCHGECT 930034 of the European Commission and the grant
A1010806 of the Academy of Sciences of the Czech Republic. The
authors acknowledge fruitfull discussions with Prof. Elias
Burnstein and Marvin L. Cohen.

\end{document}